\documentclass[12pt]{article}
\usepackage{jhep-mod}
\usepackage{bm}
\usepackage{amssymb,amsmath,amsthm}
\usepackage{mathrsfs}
\usepackage[utf8]{inputenc}
\usepackage{enumerate}
\usepackage{appendix}
\usepackage{graphicx}
\usepackage{dcolumn} 
\usepackage{bm}
\usepackage{multirow}
\usepackage{float}
\usepackage{tikz}
\usepackage{setspace}
\definecolor{purple}{rgb}{1,0,1}
\definecolor{lime}{HTML}{A6CE39} % needs xcolor
\definecolor{lime}{HTML}{A6CE39}
\newcommand{\orcidicon}{%
	\begin{tikzpicture}
	\draw[lime, fill=lime] (0,0) 
		circle [radius=0.16] 
		node[white] {{\fontfamily{qag}\selectfont \tiny ID}};
	\draw[white, fill=white] (-0.0625,0.095) 
		circle [radius=0.007];
	\end{tikzpicture}
	\hspace{-5mm}
}
\newcommand\orcidPrado{{\href{https://orcid.org/0000-0001-8073-4896}{\orcidicon}}}
\newcommand\orcidMatt{{\href{https://orcid.org/0000-0003-1088-6485}{\orcidicon}}}
\begin{document}

%========================================================
\title{\huge \leftline{Hawking--Ellis type III spacetime geometry}}

\author{\Large Prado Mart\'{\i}n-Moruno$^1$\orcidPrado{}{\sf and} 
Matt Visser$^2$\orcidMatt{}}
%%%%%%%%%%%%%%%%%%%%%%%%%%%%%%%%%%%%%%
\affiliation{$^1$ Departamento de F\'isica Te\'orica and UPARCOS,  Universidad Complutense de Madrid, \\ \null\quad E-28040 Madrid, Spain}
%%%%%%%%%%%%%%%%%%%%%%%%%%%%%%%%%%%%%%
\emailAdd{pradomm@ucm.es}
%%%%%%%%%%%%%%%%%%%%%%%%%%%%%%%%%%%%%%
\affiliation{$^2$ School of Mathematics and Statistics, Victoria University of Wellington, \\ \null\quad PO Box 600, Wellington 6140, New Zealand}
%%%%%%%%%%%%%%%%%%%%%%%%%%%%%%%%%%%%%%
\emailAdd{matt.visser@sms.vuw.ac.nz}
%%%%%%%%%%%%%%%%%%%%%%%%%%%%%%%%%%%%%
\abstract{
\parindent0pt
\parskip7pt

The type III (and the ``essential core'' type III$_0$) stress-energy tensors in the Hawking--Ellis  (Segre--Pleba\'nski) classification stand out in that there is to date no known source (either classical or semi-classical) leading to type III stress-energy. (In contrast the Hawking--Ells types I and II occur classically, and type IV is known to occur semi-classically). 
We instead start by asking the obverse question: What sort of spacetime (assuming the Einstein equations) needs a type III stress-energy to support it? One key observation is that type III is incompatible with either planar or spherical symmetry, so one should be looking at spacetimes of low symmetry (or no symmetry).
Finding such a type III spacetime is a matter of somehow finding an appropriate \emph{ansatz} for the metric, calculating the Einstein tensor, and analyzing the pattern of (Lorentz invariant) eigenvalues and eigenvectors.
Herein we report some (partial) success along these lines --- we explicitly exhibit several (somewhat unnatural) spacetime geometries with a type III Einstein tensor. 
We then build an explicit but somewhat odd Lagrangian model leading (in Minkowski space) to type III stress-energy.
While we still have no  fully acceptable general physical model for type III stress-energy, we can at least say something about what such a stress-energy tensor would entail.

\medskip
{\sc Date:} 6 June 2018; \LaTeX-ed \today

%\medskip
%{\sc arXiv:} 1805.nnnnn

\medskip
{\sc Keywords:} \\
type III; type III$_0$; stress-energy classification; Hawking--Ellis; Segre--Pleba\'nski. 
}
%=====================================================
\maketitle
%=====================================================
\def\d{{\mathrm{d}}}
\def\tr{{\mathrm{tr}}}
\parindent0pt
\parskip7pt
\clearpage

%=====================================================
\section{Introduction}

%====================================================

The Hawking--Ellis  (Segre--Pleba\'nski) classification of possible (Lorentzian signature) stress-energy tensors~\cite{Hell,Segre,Plebanski:1964} allows us to treat matter in a gravitational context without introducing specific hypotheses. Therefore, it is an essential tool in analyzing the implications of the Einstein field equations in a largely model  independent manner. Whereas the Hawking--Ellis types~I and~II have quite standard classical sources, and type IV is easily sourced by semi-classical stress-energy tensors, the type III stress-energy stands out in that there is to date no known physical source leading to such a stress-energy~\cite{Martin-Moruno:2018,Martin-Moruno:2017,LNP}. The other way that type III stands out is that it is low symmetry --- type III is incompatible with either planar or spherical symmetry --- it requires a minimum of (2+1) dimensions to even define type III~\cite{Hell, Segre,Plebanski:1964, Martin-Moruno:2018,Martin-Moruno:2017,LNP}. In contrast types I, II, and IV can all be defined in (1+1) dimensions, which certainly is compatible with either planar or spherical symmetry.

Given the oddities exhibited by the type III stress-energy tensors, one could reasonably wonder why one might be interested in such questions.
One of the reasons is the way in which the Hawking--Ellis classification interplays with and interacts with all of the classical~\cite{Hell,wormholes}, semi-classical~\cite{wormholes,Martin-Moruno:2013a,Martin-Moruno:2013b,Martin-Moruno:2015,Visser:1994,Visser:1996a,Visser:1996b,Visser:1997}, and quantum~\cite{Balakrishnan:2017,Akers:2017,Fu:2017,Fu:2016,Koeller:2015,Bousso:2015} energy conditions  --- which are in turn used as the basis for topological censorship and singularity theorems. There are also strong connections to the classification scheme built on variants of the Rainich approach, see references~\cite{Martin-Moruno:2017} and~\cite{1925,Misner:1957,Witten:1959, Senovilla:2000, Bergqvist:2001,Senovilla:2002,Bergqvist:2004,
Plebanski:1994,Torre:2013, Krongos:2015,Santos:2016,Balfagon:2007}. Due to these observations, it is of crucial importance to understand if matter described by those tensors can exists in nature and, in that case, which are its gravitational consequences.

In this article, in the first place, instead of directly looking for a physical implementation of type~III stress-energy, 
we shall ask the obverse question and look for spacetimes with a type III Einstein tensor~\footnote{As an aside we mention that there are also Petrov type III spacetimes, based on an eigenvector decomposition of the Riemann tensor rather than the stress tensor~\cite{Petrov}. However, these are quite different and mutually orthogonal classification schemes which have little to nothing to do with each other.} --- assuming the Einstein equations such a spacetime would need type III stress energy as its source. 
We then secondly develop a somewhat odd Lagrangian suitable for supporting a Hawking--Ellis type III stress-energy.
Throughout this work we will focus on a particular kind of type~III stress-energy tensor, that given by its ``essential core''~\cite{Martin-Moruno:2018}. This ``simplified" tensor is obtained by subtracting special cases of type I to simplify the (Lorentz invariant) eigenvalue structure as much as possible without disturbing the eigenvector structure. So the ``essential core'', denoted type~III$_0$, captures the fundamental characteristics of type~III tensors.

This article is outlined as follows: In section \ref{sec:typeIII} we  summarize the characteristics of type III (and type III$_0$) stress energy tensors. In section \ref{sec:metrics}, we first outline the strategy we followed to obtain spacetimes sourced by type III$_0$ stress energy tensors. Then, in subsection \ref{sec:2+1}, we present a $2+1$ dimensional example and discuss it in detail, before going to three $3+1$ dimensional examples in subsections \ref{sec:3+1a}, \ref{sec:3+1b}, and \ref{sec:3+1c}. In subsection \ref{sec:sum} we extract some conclusions about those geometries. Later, in section \ref{sec:Lagrangian}, we present particular examples of Lagrangians leading to type III$_0$ stress energy tensors---with details in subsections \ref{sec:L1} and \ref{sec:L2}. We discuss our results in section \ref{sec:dis}. Furthermore, for clarity,
we include an explanatory diagram of various stress energy tensors classifications in the appendix.
%\ref{ap}.

%=====================================================
\section{Type III and type III$_0$ stress energy}\label{sec:typeIII}
%=====================================================
Recall what type III stress-energy looks like~\cite{Martin-Moruno:2018,Martin-Moruno:2017,LNP}.
Working in (3+1) dimensions, under Lorentz similarity transformations type~III can be partially diagonalized into the form~\cite{Martin-Moruno:2018,Martin-Moruno:2017,LNP}:
\begin{equation}
T^{ab} \sim_{\hbox{\tiny L}} \left[\begin{array}{ccc|c} 
\rho&f &0&0\\ f &-\rho& f &0\\  0&f& -\rho & 0\\   \hline 0&0&0&p_3\\\end{array}\right];
\qquad
T^a{}_b \sim\left[\begin{array}{ccc|c} 
-\rho&1 &0 &0\\ 0 &-\rho& 1 &0\\  0&0& -\rho & 0\\ \hline 0&0&0&p_3\\\end{array}\right].
\end{equation}
The Lorentz invariant eigenvalues, solving $\det(T^{ab}-\lambda \eta^{ab})=0$,  are $\{-\rho,-\rho,-\rho,p_3\}$.
(This is a so-called ``generalized eigenvalue problem''~\cite{HJ1,HJ2}.)
Subtracting out as much of type I as possible; simplifying the eigenvalues as much as possible while preserving the eigenvector structure; leads to what we have called type III$_0$ stress-energy~\cite{Martin-Moruno:2018}:
\begin{equation}
(T^{ab})_{\scriptsize{\mathrm{III}}_0} \sim_{\hbox{\tiny L}} \left[\begin{array}{ccc|c} 
0&f &0&0\\ f &0& f &0\\  0&f&0& 0\\   \hline 0&0&0&0\\\end{array}\right];
\qquad
(T^a{}_b)_{\scriptsize{\mathrm{III}}_0} \sim\left[\begin{array}{ccc|c} 
0&1 &0 &0\\ 0 &0& 1 &0\\  0&0& 0& 0\\ \hline 0&0&0&0\\\end{array}\right].
\end{equation}
The Lorentz invariant eigenvalues are now $\{0,0,0,0\}$. So, we now have only one eigenvalue, $\lambda=0$, although we still have two (and only two) eigenvectors (see the appendix).
This type III$_0$ stress-energy can be invariantly characterized as $[(T^a{}_b)_{\scriptsize{\mathrm{III}}_0} ]^3 = 0$, but with  $[(T^a{}_b)_{\scriptsize{\mathrm{III}}_0} ]^2 \neq 0$, so the mixed tensor is nilpotent of order 3. Alternatively one can write 
$(T^{ab})_{\scriptsize{\mathrm{III}}_0}  = f (\ell^a s^b+s^a\ell^b)$, 
where $\ell$ is a null vector, and $s$ is a spacelike vector orthogonal to $\ell$. For more details and discussion, see reference~\cite{Martin-Moruno:2018}. As this tensor is traceless, we can conclude that the corresponding geometry has a Ricci tensor of the form $R^{ab}  = \kappa f (\ell^a s^b+s^a\ell^b)$ assuming that the gravitational phenomena are described by Einstein equations.

From the above, it is clear that the minimum dimension in which type III$_0$ can exist is (2+1) dimensions, corresponding to simply dropping the identically zero row and column in the $4\times4$ matrices presented above.
\begin{equation}
(T^{ab})_{\scriptsize{\mathrm{III}}_0} \sim_{\hbox{\tiny L}} \left[\begin{array}{ccc} 
0&f &0\\ f &0& f\\  0&f&0\\\end{array}\right];
\qquad
(T^a{}_b)_{\scriptsize{\mathrm{III}}_0} \sim\left[\begin{array}{ccc} 
0&1 &0 \\ 0 &0& 1 \\  0&0& 0\\\end{array}\right].
\end{equation}
In any dimensionality, a simple diagnostic for a type III$_0$ Einstein tensor is to check $R=0$ and then verify
\begin{equation}
G^a{}_b\;G^b{}_c\neq 0, \qquad G^a{}_b\;G^b{}_c\;G^c{}_d= 0.
\end{equation}
Equivalently one could work with the Ricci tensor
\begin{equation}
R^a{}_b\;R^b{}_c\neq 0, \qquad R^a{}_b\;R^b{}_c\;R^c{}_d= 0.
\end{equation}

%
%====================================================
\section{Some metrics with Einstein tensors of Hawking--Ellis type~III}\label{sec:metrics}
%====================================================

How did we find the type III spacetime geometries reported below? Since type III is incompatible with either spherical or planar symmetry; we knew to look at spacetimes with low symmetry. In a literature search we indeed found a paper on spacetimes with \emph{no symmetry}~\cite{Koutras}, this being closely related to the VSI (vanishing scalar invariant) spacetimes~\cite{Wils,Koutras:1992,Pravda:2002,Page:2009}. Those particular examples were not good enough for current purposes, so we kept looking. 
We also knew that the essential core tensors~\cite{Martin-Moruno:2018} are traceless, implying that the general relativistic geometrical equivalent is  Ricci-scalar-flat, $R=0$; thereby \emph{suggesting} that some mutilation\footnote{Note that that $pp$ geometries can have both lower symmetry than spherical symmetry, and $R=0$. However, they also have $G_{ab}G^{bc}=0$, thus corresponding to a type II$_0$ stress energy tensor, (or in the vacuum case, type I). That is the reason why we need to mutilate the $pp$ geometry.} of the $pp$ spacetimes might be interesting for describing the geometry corresponding to a type~III$_0$ stress-energy. Now $pp$ spacetimes can be expressed in the Brinkmann form
\begin{equation}
\d s^2=-2\d u\d v+H(u,\,x,\,y)\, \d u^2+\d x^2+\d y^2,
\end{equation}
or in the Rosen form
\begin{equation}
\d s^2=-2\d u\d v+g_{AB}(u)\d x^A\d x^B,
\end{equation}
where $x^A={x,\,y}$ (see~\cite{Cropp:2010,Cropp:2011} for a detailed discussion of this geometry).
So we started mutilating Rosen-form gravity wave spacetimes. These \emph{heuristics}  quickly led to the four examples presented below. (It must be admitted that they are quite messy spacetime geometries.)

%==============================
\subsection{(2+1)-dimensional example}\label{sec:2+1}
%==============================
This (2+1) dimensional example, despite its ultimate simplicity, was actually the last one we found.
%==============================
\subsubsection{Cartesian Kerr--Schild form}
%==============================
 Consider the (2+1) dimensional Kerr--Schild spacetime
\begin{equation}
\d s^2 = -\d t^2 + \d x^2 + \d y^2 + x\, (y+t)\, f \, (\d t-\d y)^2.
\end{equation}
Here $f$ is an arbitrary \emph{constant} and the metric is in Kerr--Schild form.\footnote{\,
This geometry can be obtained from a $pp$ wave space in Cartesian coordinates by mutilating one dimension, changing $H(t-y,\,x)\rightarrow H(t+y,\,x)$, and choosing $H(t+y,\,x)=2fx(y+t)$. (The second step being necessary for avoiding a vacuum solution).}

A brief calculation yields both $R=0$ and
\begin{equation}
G_{ab} = \left[ \begin{array}{ccc} 0 &- f & 0\\-f&0&f\\0&f&0 \end{array}\right];
\qquad
G^{ab} = \left[ \begin{array}{ccc} 0 & f & 0\\f&0&f\\0&f&0 \end{array}\right].
\end{equation}
So this metric is naturally in type III$_0$ canonical form, without any further processing being needed. It is easy to check that
\begin{equation}
G^a{}_b\;G^b{}_c\neq 0, \qquad G^a{}_b\;G^b{}_c\;G^c{}_d= 0,
\end{equation}
and that  the Jordan normal form of the mixed Einstein tensor is
\begin{equation}
\label{E:jordan}
G^a{}_b \sim \left[ \begin{array}{ccc} 0 & 1 & 0\\0&0&1\\0&0&0 \end{array}\right].
\end{equation}
The null eigenvector of $G^a{}_b$ is $\ell^a = (1,0,1)^a$; so that $\ell_a = (-1,0,1)_a$, 
while the spacelike generalized eigenvector is $s^a = (0,1,0)^a$; so that $s_a=(0,1,0)_a$.

Explicitly
\begin{equation}
g_{ab} = \eta_{ab} + x\, (y+t)\, f\,\ell_a \ell_b; \qquad
g^{ab} = \eta^{ab} + x\, (y+t)\, f\,\ell^a \ell^b.
\end{equation}
It is easy to check that
\begin{equation}
\nabla_a \ell^b = x f \, \ell_a \ell^b; \qquad \nabla_a s^b = {1\over2} (t+y) f \,\ell_a \ell^b.
\end{equation}
In particular the integral curves of any constant linear combination of $\ell^a$ and $s^a$ are geodesic.
Perhaps unexpectedly, there is a Killing vector $K = \partial_t-\partial_y$;  so we have $K^a=(1,0,-1)$.
Then $K^a\ell_a = -2$ so
\begin{equation}
K_a = -(1,0,1)_a + 2  x\, (y+t)\,f\,  (1,0,-1)_a,
\end{equation}
 and 
\begin{equation}
g_{ab}\, K^a K^b = x\, (y+t)\, (\ell_a K^a)^2\,f = 4   x\, (y+t)\,f. 
\end{equation}
Consequently the norm of the  Killing vanishes at both $x=0$ and $y+t=0$. 

The hypersurface at $x=0$ has induced 2-metric $(\d s_2)^2 = -\d t^2 + \d y^2$, and so is a (comparatively uninteresting) timelike hypersurface. In contrast the hypersurface $\Sigma$ defined by the condition $t+y=0$ has a singular induced 2-metric $(\d s_2)^2 =  \d x^2$, and so is a null hypersurface. The normal to this null hypersurface $\Sigma$ is proportional to $n_a \propto \nabla(t+y) = (1,0,1)_a$. Furthermore on this hypersurface $\Sigma$ the Killing vector reduces to $K_a \to (K_\Sigma)_a = -(1,0,1)_a \propto n_a$; so the Killing vector is normal to the hypersurface---this demonstrates that the hypersurface $\Sigma$ defined by $t+y=0$ is a Killing horizon.
Indeed the Killing vector is generally not hypersurface orthogonal. The Frobenius theorem asserts hypersurface orthogonality, $K = \alpha\, \d \beta$, if and only if $K\wedge\d K = 0$. But explicit calculation yields
\begin{equation}
K_{[a} K_{b,c]} =  -{2 (t+y)f\over3} \, \epsilon_{abc},
\end{equation}
which is in general non-zero.
So the Killing vector becomes hypersurface orthogonal only and specifically at the hypersurface $\Sigma$.
By considering the general quantity
\begin{equation}
K^a \nabla_a K^b = 2xf \,K^b + 4 x^2 (t+y) f^2\, \ell^b - 2 (t+y) f \, s^b
\end{equation}
and noting that on the $t+y=0$ Killing horizon this reduces to
\begin{equation}
K^a \nabla_a K^b \to 2 x f K^b = \kappa K^b,
\end{equation}
we see that the surface gravity of the $t+y=0$ Killing horizon is $\kappa = 2 f x$.
Note, following the construction given in Wald~\cite{Wald}, that this is compatible with 
\begin{equation}
K^{[a,b]} \, K_{[a,b]} = - 8 f^2 x^2 = -2\kappa^2.
\end{equation}

(In contrast at the timelike hypersurface $x=0$ the Killing vector is null but not hypersurface orthogonal, indeed $K^a \nabla_a K^b \to  - 2 (t+y) f \, s^b$ there,  and no meaningful definition of surface gravity can be formulated there.)

The Riemann tensor is very simple
\begin{equation}
R_{txty}= - f = R_{tyxy},
\end{equation}
other components (not related by symmetry) vanish. 
Also
\begin{equation}
R^{txty}= - f;\qquad  R^{tyxy}=f,
\end{equation}
other components (not related by symmetry) vanish. 

The Weyl tensor is zero (which is automatic in 3 dimensions). Conformal flatness in 3 dimensions is instead related to the vanishing of the Cotton tensor. Since $R=0$ 
the (3-index) Cotton tensor simplifies to $C_{abc} = 2 R_{a[b;c]}$ and explicit computation yields 
\begin{equation}
C_{abc} =  2 R_{a[b;c]} = - f^2 \,x\, \ell_a ( \ell_b s_c - \ell_c s_b).
\end{equation}
The 2-index Cotton-York tensor is 
\begin{equation}
C_{ab} = C_{acd} \; \epsilon_b{}^{cd} = + 2 f^2 x\, \ell_a \ell_b.
\end{equation}
All of the scalar invariants vanish: 
\begin{equation}
R_{abcd} R^{abcd} = R_{ab} R^{ab} = R^2 = C_{ab} C^{ab} = C_{ab} R^{ab} = 0,
\end{equation}
so this is indeed a VSI (vanishing scalar invariants) spacetime.

\clearpage
Other notable features of this (2+1) spacetime are that:
\begin{itemize}
\item The eigenvalues of the metric $g_{ab}$, with respect to the background metric $\eta_{ab}$, are 
\begin{equation}
\{-\sqrt{1+f^2x^2(t+y)^2}+fx(t+y),\,\sqrt{1+f^2x^2(t+y)^2}+fx(t+y),\,1\}. 
\end{equation}
So this metric has Lorentzian signature in the whole domain of $t,\,x,\,y$.
\item $g_{tt}=-1+x\, (y+t)\, f $, so $\d t$ is spacelike for $x\, (y+t)\, f > 1$. 
\item $g_{yy}=1+x\, (y+t)\, f $, so $\d y$ is timelike for $x\, (y+t)\, f <- 1$. 
\item
The (2+1) light cones determined by the nonsingular (2+1) metric evaluated at the timelike 2-plane $x=0$, (not those determined by the induced 2-metric on the timelike 2-plane), are the usual ones.
\item
The (2+1) light cones determined by the nonsingular (2+1) metic evaluated at the null 2-plane $t+y=0$,  (not related to the singular induced 2-metric on the null 2-plane), are the usual ones.
\end{itemize}
The causality properties seem unusual, but not entirely pathological.

%==============================
\subsubsection{Double-null coordinate form}
%==============================
The (2+1) example looks perhaps a little simpler in double-null coordinates.
Consider coordinates $u={1\over\sqrt{2}}(t-y)$ and $v={1\over\sqrt{2}}(t+y)$ so $-\d t^2+\d x^2 = - 2\d u \d v$. Then 
\begin{equation}
\d s^2 = -2\d u\d v +\d x^2+ 2 \sqrt{2} \,x v \,f \,\d u^2.
\end{equation}

Here $f$ is an arbitrary constant, the metric is in Kerr--Schild form,  and
ordering the coordinates as $(u,v,x)$ a brief calculation yields
\begin{equation}
G_{ab} = -\sqrt{2}\left[ \begin{array}{ccc} 0 & 0 & f\\0 &0&0\\f&0&0 \end{array}\right];
\qquad
G^{ab} = +\sqrt{2}\left[ \begin{array}{ccc} 0 & 0 & 0\\0&0&f\\0&f&0 \end{array}\right];
\qquad
G^a{}_b = \sqrt{2} \left[ \begin{array}{ccc} 0 & 0 & 0\\0&0&+f\\-f&0&0 \end{array}\right].
\end{equation}
This is type III$_0$ (double-null) form. (In Hawking--Ellis, and in our previous papers, we have presented the explicit matrix forms of the classification in an orthonormal basis; if we use a double null basis then things are slightly different as presented above.)

What certainly must stay the same for type III$_0$, even in double-null form, are the (easily verified) statements that $G^a{}_b\;G^b{}_c\neq 0$, and $G^a{}_b\;G^b{}_c\;G^c{}_d= 0$,
and that the Jordan normal form of the mixed Einstein tensor is still that of equation (\ref{E:jordan}).
The null eigenvector is now $\ell^a = (0,1,0)^a$; so that $\ell_a =-(1,0,0)_a$. 
The spacelike generalized eigenvector is now $s^a=(0,0,1)^a$, so that $s_a=(0,0,1)^a$.
The Riemann tensor is now (if anything) even simpler
\begin{equation}
R_{uvux}= -\sqrt{2} f; \qquad\qquad R^{uvvx}= -\sqrt{2} f.
\end{equation}
\enlargethispage{20pt}
Other components (not related by symmetry) vanish. 
The Killing vector becomes $\partial_u$;  that is $K^a=(1,0,0)$,
whereas $K_a = \left(+2\sqrt{2} xvf,-1,0\right)_a$ and $g_{ab} \, K^a K^b = 2\sqrt{2}xvf$. 
The Killing horizon $\Sigma$ is now specified by the hypersurface $v=0$.  (As previously, the hypersurface $x=0$ has induced 2-metric $(\d s_2)^2 = -2 \d u \d v$, and so is again timelike.)
Other tensorial properties of the spacetime carry over without modification.

One (trivial) way of going from (2+1) to (3+1) dimensions is by simply adding on an extra flat dimension --- this is not particularly interesting ---  a less trivial construction involves ``distorting'' in the extra dimension.

%==============================
\subsection{(3+1)-dimensional examples}\label{sec:3+1}
%==============================

We now present three examples of type III spacetime geometry in (3+1) dimensions.

%==============================
\subsubsection{First and simplest (3+1)-dimensional example}\label{sec:3+1a}
%==============================

Let us consider a $pp$ spacetime in the Rosen form
\begin{equation}
\d s^2=-2\d u\d v+g_{AB}(u)\d x^A\d x^B.
\end{equation}
Here $x^A=\{x,\,y\}$, and $u$ and $v$ are null coordinates. This spacetime is sourced by a type II$_0$ stress energy tensor, that is $G_{ab}\,G^{bc}=0$. In order to avoid this (for our purposes trivial) conclusion, we \emph{mutilate} the space by making the change $u\longleftrightarrow y$.  A particularly simple example is the following geometry (which was actually the first we found)
\begin{equation}
\d s^2 = -\d t^2 + \d x^2+\d y^2+\d z^2 + 2J(y) \,\d x\, \{\d z - \d t\}.
\end{equation}
This is a \emph{mutilation} of $pp$ spacetime in the Rosen form.
Note there are three linearly independent Killing vectors $\partial_t$, $\partial_x$, and $\partial_z$.

One easily calculates the only nonzero components of the Einstein tensor
\begin{equation}
G_{xz} = - G_{xt} = - {1\over 2} J_{,yy} ,
\qquad
\hbox{and}
\qquad
G_{zz} = - G_{zt} = G_{tt} = + {1\over2} (J_{,y})^2.
\end{equation}
Therefore, at a minimum we need $J_{,yy}\neq0$ in order to get a type III tensor.
One easily verifies that the Ricci scalar is zero, $R = 0$. Moreover, $R_{abcd}R^{abcd}=0$ and $R^{ab}R_{ab}=0$.
If we now write 
\begin{equation}
A = -{1\over 2}  J_{,yy}; \qquad 
E=  + {1\over4} (J_{,y})^2;
\end{equation}
then
\begin{equation}
G_{ab}\, \d x^a \d x^b =  2A\,\d x(\d z-\d t) + 2E\, (\d z-\d t)(\d z-\d t).
\end{equation} 
Rearrange this
\begin{equation}
G_{ab}\, \d x^a \d x^b =  2(A\d x+E(\d z-\d t))\;(\d z-\d t).
\end{equation} 
This is now of the required type III$_0$ form,
$G_{ab} = f \,(\ell_a s_b + s_a \ell_b)$,
provided we set
\begin{equation}
\ell_a \d x^a = \d z-\d t; \qquad \hbox{and} \qquad f \; s_a \d x^a = A\d x+E(\d z-\d t).
\end{equation}
Note the 1-form $\ell = \d z-\d t$ is null with respect to both $\eta$ and $g$,
whence
\begin{equation}
f = A;  \qquad \hbox{and} \qquad s_a \d x^a = \d x+\left(E\over A\right)\; (\d z-\d t).
\end{equation}
that is
\begin{equation}
f = {1\over 2}  J_{,yy};  \qquad \hbox{and} \qquad s_a \d x^a = \d x-
 {1\over 2} \left(  (J_{,y})^2\over  J_{,yy}\right)\; (\d z-\d t).
\end{equation}
Furthermore
\begin{equation}
\ell_a = (-1,0,0,1)_a; \qquad \hbox{and} \qquad \ell^a = (1,0,0,1)^a.
\end{equation}
Similarly
\begin{equation}
s_a = \left(-{E\over A},1,0,{E\over A}\right)_a; 
\qquad \hbox{and} \qquad
s^a = \left(-J+{E\over A},1,0,-J+{E\over A}\right)^a.
\end{equation}
Note that $s$ is normalized: $s_as^a=1$. We  also have $\nabla_a \ell^b=0$ and $s^a \nabla_a s^b = 0$. 
In particular $\ell^a$ is a covariantly constant null Killing vector, and the integral curves of $s^a$ are geodesic.

\clearpage
One can also check the key features above by brute force, symbolically calculating $G^a{}_b$, (\emph{e.g.} using {\sf Maple}), and verifying that
\begin{equation}
(G^2)^a{}_b= f^2 \ell^a \ell_b = f^2 \left[ \begin{array}{cccc} -1 & 0 & 0 & 1\\0 &0&0&0\\0&0&0&0\\-1 &0&0&1 \end{array}\right]^a_{\;\;b};
\qquad \hbox{while} \qquad (G^3)^a{}_b = 0.
\end{equation}

One easily verifies that the Weyl tensor is non-zero $C_{abcd}\neq 0$, but $C_{abcd} C^{abcd} = 0$.
Also $R_{abcd} R^{abcd} = 0$ and $ G^{ac} R_{abcd} = 0=  G^{ac} C_{abcd}$.
This spacetime geometry would need to be sourced by a type III$_0$ stress-energy tensor to satisfy the Einstein equations.

%================================
\subsubsection{Second more general (3+1)-dimensional example}\label{sec:3+1b}
%================================

For a second example, consider the more complicated metric
\def\d{{\mathrm{d}}}
\begin{equation}
\d s^2 = -\d t^2 + \d x^2+\d y^2+\d z^2 
+ 2\left\{J(x,y,z-t) \d x+K(x,y,z-t)\d y\right\}\, \left\{\d z - \d t\right\}.
\end{equation}
This is not Kerr--Schild, though it is vaguely reminiscent thereof. \\
This is not a $pp$ spacetime, though it is vaguely reminiscent thereof.\\
There is now only one Killing vector, namely $\partial_t + \partial_z$, that is $K^a=(1,0,0,1)^a$.

One easily calculates the only non-zero components of the Einstein tensor
\begin{eqnarray}
G_{xz} = - G_{xt} &=& {1\over 2} [J_{,yy}-K_{,xy}] =  {1\over 2}\partial_y[J_{,y}-K_{,x}]; 
\\
G_{yz}=-G_{yt} &=& {1\over2} [K_{,xx} - J_{,xy}]= - {1\over2} \partial_x[J_{,y}- K_{,x}];
\end{eqnarray}
and 
\begin{equation}
G_{zz} = - G_{zt} = G_{tt} = -J_{,xz}-K_{,yz} - {1\over2} (J_{,y}-K_{,x})^2
=  -\partial_z[J_{,x}+K_{,y}] - {1\over2} (J_{,y}-K_{,x})^2.
\end{equation}
One also easily verifies that the Ricci scalar is zero $R = 0$. 
If we now write 
\begin{equation}
A = {1\over 2}  [J_{,yy}-K_{,xy}]; \qquad B= {1\over2} [K_{,xx} - J_{,xy}]; 
\end{equation}
and
\begin{equation}
E=  {1\over2} \left(-J_{,xz}-K_{,yz} - {1\over2} (J_{,y}-K_{,x})^2\right);
\end{equation}
then it is easy to see that
\begin{eqnarray}
G_{ab} \, \d x^a \d x^b &=&  2A[\d x(\d z-\d t)] - 2B[\d y(\d z-\d t)] 
+ 2E (\d z-\d t)^2.
\end{eqnarray} 
Rearrange this to get
\begin{equation}
G_{ab} \,\d x^a \d x^b =  2[(A\d x-B\d y)(\d z-\d t)] + 2E (dz-dt)^2.
\end{equation} 
A further rearrangement yields
\begin{equation}
G_{ab} \,\d x^a \d x^b =  2[A\d x-B\d y+E(\d z-\d t)](\d z-\d t). 
\end{equation} 
This now is of the required type III$_0$ form, $G_{ab} = f (\ell_a s_b + s_a \ell_b)$, provided we set
\begin{equation}
\ell_a \d x^a = \d z-\d t; \qquad \hbox{and} \qquad f s_a \d x^a = A\d x-B\d y+E(\d z-\d t).
\end{equation}
Note the 1-form $\ell = \d z-\d t$ is null with respect to both $\eta$ and $g$,
whence
\begin{equation}
f = \sqrt{A^2+B^2}; \qquad \hbox{and} \qquad s_a \d x^a = {A\d x-B\d y+E(\d z-\d t)\over\sqrt{A^2+B^2}}.
\end{equation}
Note
\begin{equation}
\ell_a = (-1,0,0,1)_a; \qquad \ell^a=(1,0,0,1)^a;  \qquad \nabla_a \ell^b = 0.
\end{equation}
But this means $K^a=\ell^a$; the Killing vector is everywhere null and covariantly constant.

One can also check the key features above by brute force, symbolically calculating $G^a{}_b$, (\emph{e.g.} using {\sf Maple}), and verifying that $(G^2)^a{}_b = f^2 \ell^a \ell_b$, while $(G^3)^a{}_b = 0$.
One easily verifies that the Weyl tensor is non-zero $C_{abcd}\neq 0$, but $C_{abcd} C^{abcd} = 0$.
Also $R_{abcd} R^{abcd} = 0$ and $ G^{ac} R_{abcd} = 0=  G^{ac} C_{abcd}$.
This spacetime geometry would need to be sourced by a type III$_0$ stress-energy tensor to satisfy the Einstein equations.

%======================================================
\subsubsection{Third even more general (3+1) dimensional example}\label{sec:3+1c}
%======================================================

Here is an even more general example. Consider the spacetime metric
\begin{equation}
\d s^2 = \d x^2 + \d y^2 + 2 \d u \d v + 2 \{J(x,y,u)\d x + K(x,y,u)\d y + L(x,y,u)\d u \} \d u.
\end{equation}
There is only one Killing vector, namely $\partial_v$. Ordering the coordinates as $x^a=(u,v,x,y)$ we have $K^a=(0,1,0,0)^a$.

The inverse metric is easily calculated
\begin{eqnarray}
g^{ab} \partial_a \partial_b &=& \partial_x^2 +\partial_y^2 + 2 \partial_u\partial_v 
- 2\{J(x,y,u) \partial_x + K(x,y,u) \partial_y\} \partial_v 
\nonumber\\ &&
+ \{ J(x,y,u)^2 + K(x,y,u)^2 - 2L(x,y,u)\} \partial_v^2.
\end{eqnarray}
Explicitly
\begin{equation}
g_{ab}=\left[\begin{array}{cc|cc} 
2L & 1 & J & K\\
1 &0&0&0\\
\hline
J&0&1&0\\
K&0&0&1
\end{array}\right];
\qquad
g^{ab}=\left[\begin{array}{cc|cc} 
0 & 1 & 0 & 0\\
1 &-2L+J^2+K^2&-J&-K\\
\hline
0&-J&1&0\\
0&-K&0&1
\end{array}\right].
\end{equation}

The the nonzero components of the Einstein tensor are:
\begin{equation}
G_{xu} = {1\over2} \partial_y \{ J_{,y} - K_{,x}\};
\qquad
G_{yu} = -{1\over2} \partial_x \{ J_{,y} - K_{,x}\};
\end{equation}
and
\begin{equation}
G_{uu} = + L_{,xx}+L_{,yy}- \partial_u  \{ J_{,x} + K_{,y}\} -{1\over2}  \{ J_{,y} - K_{,x}\}^2.
\end{equation}
This implies:
\begin{equation}
R=0,
\end{equation}
\begin{equation}
(G^2)_{ab}  \; \d x^a \d x^b = f^2 \d u^2,
\end{equation}
\begin{equation}
(G^3)_{ab} \;  \d x^a \d x^b =0.
\end{equation}
Indeed in terms of the 1-form $s_a \,\d x^a$ the Einstein tensor is of type III$_0$ form:
\begin{equation}
G_{ab}\,  \d x^a \d x^b = f  \, \{ (s_a \d x^a) \; \d u + \d u\; (s_a \d x^a)\},
\end{equation}
where
\begin{equation}
\ell=\d u; \qquad 
f \;s = A \d x + B \d y + E \d u.
\end{equation}
Here
\begin{equation}
A = {1\over2} \partial_y \{ J_{,y} - K_{,x}\}; \qquad 
B= -{1\over2} \partial_x \{ J_{,y} - K_{,x}\}; 
\end{equation}
while
\begin{equation}
E=  + L_{,xx}+L_{,yy}- \partial_u  \{ J_{,x} + K_{,y}\} -{1\over2}  \{ J_{,y} - K_{,x}\}^2;
\end{equation}
and
\begin{equation}
f = \sqrt{A^2 + B^2}.
\end{equation}
Note 
\begin{equation}
\ell_a = (1,0,0,0)_a; \qquad \ell^a=(0,1, 0,0)^a;  \qquad \nabla_a \ell^b = 0.
\end{equation}
But this means $K^a=\ell^a$; that is the Killing vector is everywhere null and covariantly constant.
One easily verifies that the Weyl tensor is non-zero $C_{abcd}\neq 0$, but $C_{abcd} C^{abcd} = 0$.
Also $R_{abcd} R^{abcd} = 0$ and $ G^{ac} R_{abcd} = 0=  G^{ac} C_{abcd}$.
This spacetime geometry would need to be sourced by a type III$_0$ stress-energy tensor to satisfy the Einstein equations.

%==============================
\subsection{Summary regarding type III geometries}\label{sec:sum}
%==============================

From the above we see that type III spacetime geometries we have found are somewhat odd and unusual in their properties --- with the low symmetry making them somewhat tricky to deal with. The algebraic properties of the curvature tensor (at least for the specific examples we found above) are relatively simple, whereas the causal structure is intricate. It is not entirely clear whether these four examples can be further generalized, nor is there any obvious underlying pattern. The physical interpretation of these spacetimes
is perhaps less than clear.

\enlargethispage{20pt}
The three explicit (3+1) examples all possess a covariantly constant null Killing vector --- this is one of the surviving features of the $pp$ spacetimes which we deliberately mutilated to get a type III$_0$ Ricci tensor.

%========================================================
\section{Lagrangian model  for type III stress-energy}\label{sec:Lagrangian}
%========================================================

\def\L{{\mathrm{L}}}
Based on these examples of type III spacetimes, is it now possible to \emph{guess} a suitable matter Lagrangian? And so finally get a type III stress tensor? Here we report some limited success along these lines.

%========================================================
\subsection{Lagrangian}\label{ssec:L}
%========================================================
\def\L{{\mathcal{L}}}
Recall that for type III$_0$ stress-energy we want
\begin{equation}
T_{ab} = f (\ell_a s_b + s_a \ell_b);   \qquad \ell^2 = 0; \qquad s^2 = +1; \qquad \ell\cdot s =0.
\end{equation}
Here is a suitable (if somewhat unusual) ansatz: Consider this Lagrangian, depending only on the divergence $\nabla_a A^a$ of some vector field $A^a$:
\begin{equation}
\L = F(\nabla_a A^a).
\end{equation}
Then the EOM for the vector field $A^a$ are simply
\begin{equation}
\nabla_b \left( F'(\nabla_a A^a)\right) = 0.
\end{equation}
That is
\begin{equation}
F''(\nabla_a A^a)  \;\; \nabla_b (\nabla_a A^a) = 0.
\end{equation}

As long as one is not at a critical point, $F''(\nabla_a A^a) \neq 0$, this further simplifies to
\begin{equation}
\nabla_b (\nabla_a A^a) = 0,
\end{equation}
implying 
\begin{equation}
(\nabla_a A^a) = \hbox{(constant)}.
\end{equation}
The stress-energy tensor is
\begin{equation}
T_{ab} = {1\over2} F'(\nabla_a A^a) \;\;\{ \nabla_a A_b+ \nabla_b A_a \} -  {1\over2}   F(\nabla_a A^a) g_{ab}.
\end{equation}
That is, in terms of the Lie derivative
\begin{equation}
T_{ab} = F' (\hbox{constant}) \; \mathscr{L}_A g_{ab}  -  {1\over2}   F(\hbox{constant})\, g_{ab}.
\end{equation}
%\enlargethispage{30pt}
Now consider two special cases. While ultimately we would like to self-consistently find a Lagrangian-based source for the  four type III$_0$ geometries discussed above, for now we will settle for a ``proof of principle'' by working in flat Minkowski space.

%========================================================
\subsection{Linear ansatz}\label{sec:L1}
%========================================================
In flat Minkowski space $g_{ab}\to \eta_{ab}$, and one specific solution of the vector field EOM is the linear expression
\begin{equation}
A^a = M^a{}_b \, x^b.
\end{equation}
Here $M^a{}_b$ is a constant matrix.
Then $\nabla_a A^b = M^b{}_a$ and so
\begin{equation}
T_{ab} = {1\over2} F'(M^c{}_c) \;\{ M_{ba}+ M_{ab} \} -  {1\over2}   F(M^c{}_c)\; \eta_{ab}.
\end{equation}
Quite generally this expression could algebraically yield \emph{any} of the Hawking--Ellis stress-energy types. 

\clearpage
If we now take the even more specific case $M_{ab} = s_a \ell_b$, with $s^2=+1$, $\ell^2=0$, and $\ell\cdot s = 0$, then we have $M^c{}_c=0$ and
\begin{equation}
T_{ab} = {1\over2} F'(0) \;\;\{ s_a \ell_b + \ell_a s_b \} -  {1\over2}   F(0) \; \eta_{ab}.
\end{equation}
Finally choosing $F'(0)=2f \neq 0$ this yields a type III stress-energy in Minkowski space.
(Furthermore the sub-case $F(0)=0$ yields a type III$_0$ stress-energy.)%
\footnote{
In a similar vein, the choice $M_{ab}= \ell_a \ell_b$ with $\ell^2=0$ would yield type II, specializing to type II$_0$ if $F(0)=0$. Choosing $M_{ab}$ to be a generic diagonal matrix yields (generic) type I stress-energy. Finally, choosing $M_{ab} = s_a V_b$ with $s^2=+1$, $V^2=-1$, and $V\cdot s = 0$ yields type IV, specializing to type IV$_0$ if $F(0)=0$.
}
%========================================================
\subsection{Wave ansatz}\label{sec:L2}
%========================================================

Again work in flat Minkowski space $g_{ab}\to \eta_{ab}$, but now consider the plane wave ansatz
\begin{equation}
A^a = s^a \sin(k_c x^c).
\end{equation}
Then $\nabla_b A^a = k_b \,s^a \cos(k_c x^c)$ and so  $\nabla_a A^a = k_a \,s^a \cos(k_c x^c)$. Because of the oscillating cosine factor the vector field EOM is satisfied only if $s^a k_a = 0$, in which case $\nabla_a A^a=0$. Then we have
\begin{equation}
T_{ab} = {1\over2} F'(0) \;\{ s_a k_b+ k_a s_b \} \cos(k_c x^c) -  {1\over2}   F(0)  \; \eta_{ab}.
\end{equation}
If we now choose $s^2=+1$ and $k\to\ell$ with $\ell^2=0$, and $\ell\cdot s = 0$ then
\begin{equation}
T_{ab} = {1\over2} F'(0) \;\{ s_a \ell_b+ \ell_a s_b \} \cos(k_c x^c) -  {1\over2}   F(0)  \; \eta_{ab}.
\end{equation}
This is of type III whenever the cosine is non-zero. (And furthermore is of type III$_0$ whenever $F(0)=0$ and the cosine is nonzero.)%
\footnote{
In a similar vein, the choice $s_a=k_a=\ell_a$ with $\ell^2=0$ would yield type II, specializing to type II$_0$ if $F(0)=0$. Furthermore choosing $s^2=+1$, $k^2=-1$, and $k\cdot s = 0$ yields type IV, specializing to type IV$_0$ if $F(0)=0$. Finally type I is the generic case when there is no particular relationship between the vectors $s$ and $k$.}

\clearpage
%==============================
\subsection{Summary regarding type III stress-energy}\label{sec:Lsum}
%==============================

It must be said that the Lagrangian $\L = F(\nabla_a A^a)$ is a very unusual Lagrangian that does not conform to any of the standard versions of ``matter'' occurring (for instance) in the standard model of particle physics.
That we can get a ``proof of principle'' type~III stress-energy when working on Minkowski space is encouraging; that so little can be said in general curved spacetimes is discouraging.

%========================================================
\section{Discussion}\label{sec:dis}
%========================================================

In the  Hawking--Ellis (Segre--Pleba\'nski) classification the type III stress-energy tensor (and the closely related type III$_0$ essential core) stand out in that there is relatively little understanding of this class of stress-energy tensors and (assuming the Einstein equations) the related spacetimes. 
In this article we have first given 4 specific examples of spacetimes supported by type III$_0$ stress-energy, and have then developed ``proof of principle'' Lagrangian models for type III stress-energy in Minkowski space.

Unfortunately we have not yet been able to ``close the loop'', to self-consistently determine a type III spacetime geometry supported by an explicit Lagrangian-based type III matter source. 
This remaining part of the problem looks rather difficult --- the type III  spacetime geometries are all rather unusual, and the only Lagrangian model we have for type III  stress-energy is if anything, extremely odd.
Nevertheless, the results reported in this article represent real progress compared to what was previously known. At this point we can already suggest that a robust assumption when investigating general relativistic spacetimes is to focus attention on type I, II and IV stress energy tensors.

%========================================================
\section*{Acknowledgments}
%========================================================
PMM acknowledges financial support from the project FIS2016-78859-P (AEI/FEDER, UE), and financial support provided through the Research Award L'Or\'eal-UNESCO FWIS (XII Spanish edition).
MV acknowledges financial support via the Marsden Fund administered by the Royal Society of New Zealand.

\appendix
%\addappheadtotoc
%\appendixpage
%%%%%%%%%%%%%%%%%%%%%%%%%%%%%%%%%%%%%%%%

%=============================================================
\section*{Appendix: Graphical classification of stress-energy tensors}
\label{ap}
\begin{figure}[htbp!]
\begin{center}
\includegraphics[scale=0.50]{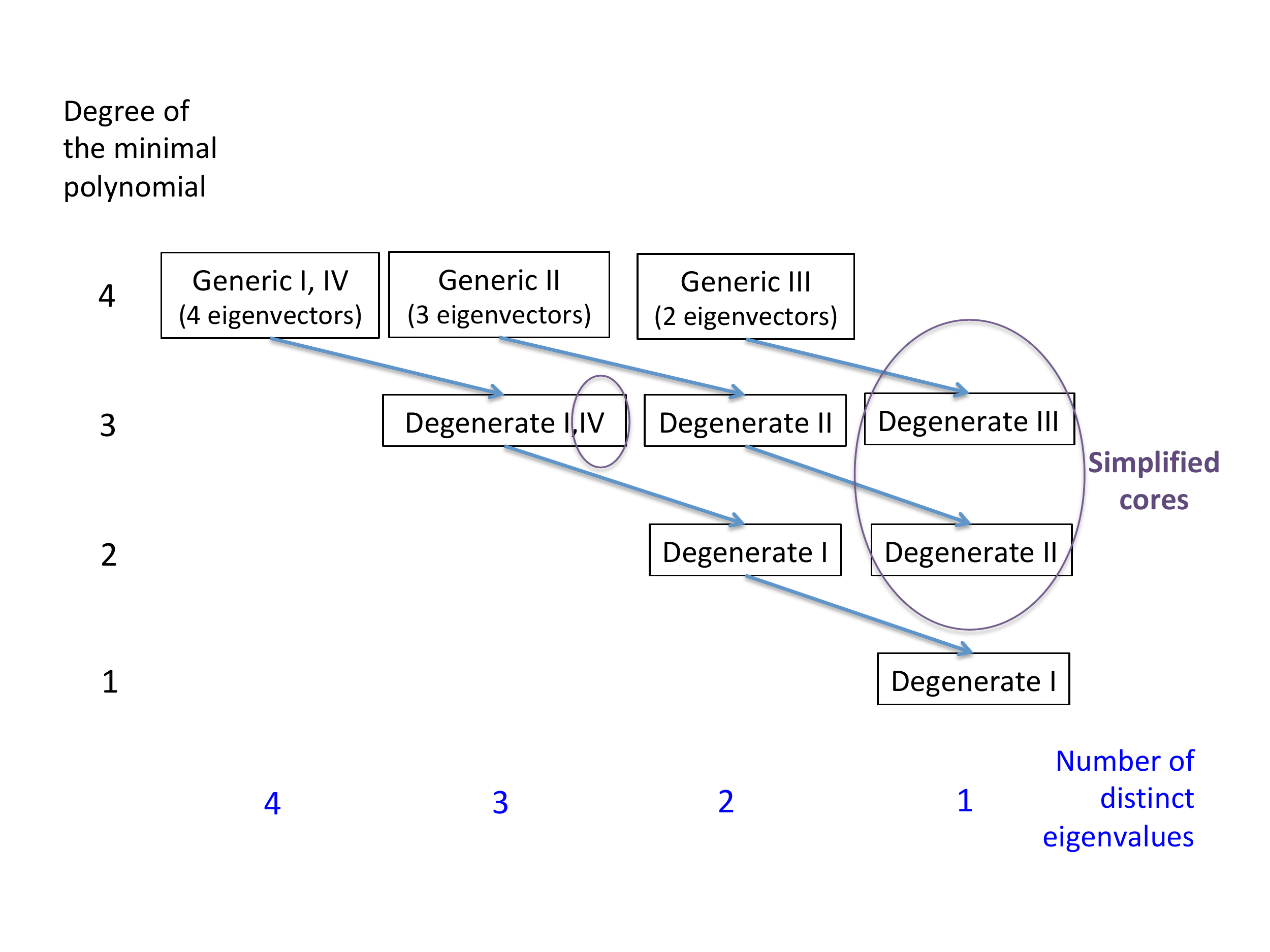}
\caption{In this diagram we show the relation between different stress energy tensors classifications in $1+3$ dimensions. Generic type I and IV, both corresponding to degree 4.4, are stable under perturbations, whereas the other will decay into those. Degenerate types have the same number of eigenvectors as the generic cases, but fewer distinct eigenvectors \cite{Martin-Moruno:2017}.}
\label{SETclas}
\end{center}
\end{figure}

\addcontentsline{toc}{section}{Appendix: Graphical classification of stress-energy tensors}
%=============================================================
Let us very briefly review, in a graphical manner, the different possible classifications of the stress energy tensor. In the diagram depicted in figure \ref{SETclas} we show the relation between the Hawking--Ellis (Segre--Pleba\'nski) classification, which is based on the eigenvectors of the stress energy tensor, compared with the Rainich-like classification based on the degree of the minimal polynomials, related with the multiplicity of the eigenvalues of the mixed tensor $T^a{}_b$. The simplified cores are the special cases where all real eigenvalues vanish, implying (but stronger than) the condition that the stress-energy is traceless.

\vfill
%========================================================
%========================================================

%========================================================

\begin{thebibliography}{69}  
%========================================================
%========================================================

\bibitem{Hell} S. W. Hawking and G. F. R. Ellis,
\emph{The large scale structure of space-time},\\
(Cambridge University Press, Cambridge, 1973).
%========================================================

\bibitem{Segre}
Corrado Segre,  ``Sulla teoria e sulla classificazione delle omografie in uno spazio lineare ad uno numero qualunque di dimensioni'', \\
Memorie della R. Accademia dei Lincei {\bf3a} (1884) 127.
%========================================================

\bibitem{Plebanski:1964}
Jerzy Pleba\'nski, ``The algebraic structure of the tensor of matter", \\
Acta Physica Polonica. {\bf26} (1964) 963.
%========================================================

\bibitem{Martin-Moruno:2018}
  P.~Mart\'in--Moruno and M.~Visser,
  ``Essential core of the Hawking--Ellis types'',\\
  Classical and Quantum Gravity {\bf 35} (2018) no.12, 125003\\
  doi:10.1088/1361-6382/aac147
  [arXiv:1802.00865 [gr-qc]].
  
 %========================================================
 
 \bibitem{Martin-Moruno:2017}
  P.~Mart\'in--Moruno and M.~Visser,
  ``Generalized Rainich conditions, generalized stress-energy conditions, and the Hawking--Ellis classification'',\\
  Classical and Quantum Gravity {\bf 34} (2017) no.22,  225014\\
  doi:10.1088/1361-6382/aa9039
  [arXiv:1707.04172 [gr-qc]].
  %%CITATION = doi:10.1088/1361-6382/aa9039;%%
 %========================================================
\bibitem{LNP}
  P.~Mart\'in--Moruno and M.~Visser,
  ``Classical and semi-classical energy conditions'',\\
  Fundam.\ Theor.\ Phys.\  {\bf 189} (2017) 193 (Lecture Notes in Physics)\\
  doi:10.1007/978-3-319-55182-1\_9
  [arXiv:1702.05915 [gr-qc]].
  %%CITATION = doi:10.1007/978-3-319-55182-1_9;%%
  %2 citations counted in INSPIRE as of 30 Jun 2017
  
  
  
%========================================================
% CLASSICAL ENERGY CONDITIONS
%========================================================
  
  
%========================================================
% SEMI-CLASSICAL ENERGY CONDITIONS
%========================================================
  
  \bibitem{wormholes}
M.~Visser,
  \emph{Lorentzian wormholes: From Einstein to Hawking},\\
  (AIP press, now Springer--Verlag, New York, 1995) %412 p
  %111 citations counted in INSPIRE as of 30 Jan 2018
  
  
 %========================================================

\bibitem{Martin-Moruno:2013a} 
  P.~Mart\'in--Moruno and M.~Visser,
  ``Classical and quantum flux energy conditions for quantum vacuum states'',
  Phys.\ Rev.\ D {\bf 88}, no. 6, 061701 (2013)
  doi:10.1103/PhysRevD.88.061701
  [arXiv:1305.1993 [gr-qc]].
  %%CITATION = doi:10.1103/PhysRevD.88.061701;%%
  %18 citations counted in INSPIRE as of 30 Jun 2017
  %========================================================

\bibitem{Martin-Moruno:2013b}
  P.~Mart\'in--Moruno and M.~Visser,
  ``Semiclassical energy conditions for quantum vacuum states'',
  JHEP {\bf 1309} (2013) 050
  doi:10.1007/JHEP09(2013)050
  [arXiv:1306.2076 [gr-qc]].
  %%CITATION = doi:10.1007/JHEP09(2013)050;%%
  %18 citations counted in INSPIRE as of 30 Jun 2017
%========================================================
  
  \bibitem{Martin-Moruno:2015}
  P.~Mart\'in--Moruno and M.~Visser,
  ``Semi-classical and nonlinear energy conditions'', 
  Proceedings of The Fourteenth Marcel Grossmann Meeting, 1442-1447. World Scientific 2017.
  doi:10.1142/9789813226609\_0126 
  [arXiv:1510.00158 [gr-qc]].
  %%CITATION = ARXIV:1510.00158;%%
  %3 citations counted in INSPIRE as of 30 Jun 2017
%========================================================

%========================================================
  
  \bibitem{Visser:1994}
  M.~Visser,
  ``Scale anomalies imply violation of the averaged null energy condition'',\\
  Phys.\ Lett.\ B {\bf 349} (1995) 443
  doi:10.1016/0370-2693(95)00303-3
  [gr-qc/9409043].
  %%CITATION = doi:10.1016/0370-2693(95)00303-3;%%
  %49 citations counted in INSPIRE as of 13 Jul 2017
 %========================================================
 \enlargethispage{40pt}
 \bibitem{Visser:1996a}
  M.~Visser,
  ``Gravitational vacuum polarization. 1: Energy conditions in the Hartle-Hawking vacuum'',
  Phys.\ Rev.\ D {\bf 54} (1996) 5103
  doi:10.1103/PhysRevD.54.5103
  [gr-qc/9604007].
  %%CITATION = doi:10.1103/PhysRevD.54.5103;%%
  %58 citations counted in INSPIRE as of 30 Jan 2018
  
  \bibitem{Visser:1996b}
  M.~Visser,
  ``Gravitational vacuum polarization. 2: Energy conditions in the Boulware vacuum'',
  Phys.\ Rev.\ D {\bf 54} (1996) 5116
  doi:10.1103/PhysRevD.54.5116
  [gr-qc/9604008].
  %%CITATION = doi:10.1103/PhysRevD.54.5116;%%
  %68 citations counted in INSPIRE as of 30 Jan 2018
  
\bibitem{Visser:1997}
  M.~Visser,
  ``Gravitational vacuum polarization. 4: Energy conditions in the Unruh vacuum'',
  Phys.\ Rev.\ D {\bf 56} (1997) 936
  doi:10.1103/PhysRevD.56.936
  [gr-qc/9703001].
  %%CITATION = doi:10.1103/PhysRevD.56.936;%%
  %49 citations counted in INSPIRE as of 13 Jul 2017
  %========================================================
  

 
%========================================================
% QUANTUM ENERGY CONDITIONS
%========================================================
  
   
\bibitem{Balakrishnan:2017} 
  S.~Balakrishnan, T.~Faulkner, Z.~U.~Khandker and H.~Wang,
  ``A General Proof of the Quantum Null Energy Condition'',
  arXiv:1706.09432 [hep-th].
  %%CITATION = ARXIV:1706.09432;%%
%========================================================

\bibitem{Akers:2017}
  C.~Akers, V.~Chandrasekaran, S.~Leichenauer, A.~Levine and A.~Shahbazi Moghaddam,\\
  ``The Quantum Null Energy Condition, Entanglement Wedge Nesting, and Quantum Focusing'',
  arXiv:1706.04183 [hep-th].
  %%CITATION = ARXIV:1706.04183;%%
  %1 citations counted in INSPIRE as of 20 Jul 2017
  
 %========================================================


\bibitem{Fu:2017} 
  Z.~Fu, J.~Koeller and D.~Marolf,
  ``The Quantum Null Energy Condition in Curved Space'',
  arXiv:1706.01572 [hep-th].
  %%CITATION = ARXIV:1706.01572;%%
  %2 citations counted in INSPIRE as of 30 Jun 2017
%\cite{Fu:2016avb}
%========================================================

\bibitem{Fu:2016} 
  Z.~Fu and D.~Marolf,
  ``Does horizon entropy satisfy a Quantum Null Energy Conjecture?'',\\
  Classical and Quantum Gravity {\bf 33}, no. 24, 245011 (2016)
  doi:10.1088/0264-9381/33/24/245011\\{}
  [arXiv:1606.04713 [hep-th]].
  %%CITATION = doi:10.1088/0264-9381/33/24/245011;%%
  %4 citations counted in INSPIRE as of 30 Jun 2017
%========================================================

\bibitem{Koeller:2015} 
  J.~Koeller and S.~Leichenauer,
  ``Holographic Proof of the Quantum Null Energy Condition'',\\
  Phys.\ Rev.\ D {\bf 94}, no. 2, 024026 (2016)
  doi:10.1103/PhysRevD.94.024026\\{}
  [arXiv:1512.06109 [hep-th]].
  %%CITATION = doi:10.1103/PhysRevD.94.024026;%%
  %22 citations counted in INSPIRE as of 30 Jun 2017
%\cite{Bousso:2015wca}
%========================================================

\bibitem{Bousso:2015} 
  R.~Bousso, Z.~Fisher, J.~Koeller, S.~Leichenauer and A.~C.~Wall,\\
  ``Proof of the Quantum Null Energy Condition'',
  Phys.\ Rev.\ D {\bf 93}, no. 2, 024017 (2016)\\
  doi:10.1103/PhysRevD.93.024017
  [arXiv:1509.02542 [hep-th]].
  %%CITATION = doi:10.1103/PhysRevD.93.024017;%%
  %34 citations counted in INSPIRE as of 30 Jun 2017
%========================================================
  
%========================================================
%  RAINICH
%========================================================


\bibitem{1925}
G.~Y.~Rainich, ``Electrodynamics in the general relativity theory'',\\
 Trans. Am. Math. Soc. {\bf27} (1925) 106.
%========================================================

\bibitem{Misner:1957}
  C.~W.~Misner and J.~A.~Wheeler,
  ``Classical physics as geometry: Gravitation, electromagnetism, unquantized charge, and mass as properties of curved empty space'',
  Annals Phys.\  {\bf 2} (1957) 525.
  doi:10.1016/0003-4916(57)90049-0
  %%CITATION = doi:10.1016/0003-4916(57)90049-0;%%
  %351 citations counted in INSPIRE as of 30 Jun 2017
 %========================================================
  
  \bibitem{Witten:1959}
  L.~Witten,
  ``Geometry of gravitation and electromagnetism'',
  Phys.\ Rev.\  {\bf 115} (1959) 206.
  doi:10.1103/PhysRev.115.206
  %%CITATION = doi:10.1103/PhysRev.115.206;%%
  %8 citations counted in INSPIRE as of 30 Jun 2017
%========================================================


\bibitem{Senovilla:2000}
  J.~M.~M.~Senovilla,
  ``General electric magnetic decomposition of fields, positivity and Rainich-like conditions'',
  gr-qc/0010095.
  %%CITATION = GR-QC/0010095;%%
  %10 citations counted in INSPIRE as of 30 Jun 2017
%========================================================

\bibitem{Bergqvist:2001}
   G.~Bergqvist and J.~M.~M.~Senovilla,
  ``Null cone preserving maps, causal tensors and algebraic Rainich theory'',
  Classical and Quantum Gravity {\bf 18} (2001) 5299
  doi:10.1088/0264-9381/18/23/323
  [gr-qc/0104090].
  %%CITATION = doi:10.1088/0264-9381/18/23/323;%%
  %30 citations counted in INSPIRE as of 30 Jun 2017
%========================================================
  
\bibitem{Senovilla:2002}
  J.~M.~M.~Senovilla,
  ``Superenergy tensors and their applications'',
  math-ph/0202029.
  %%CITATION = MATH-PH/0202029;%%
  %1 citations counted in INSPIRE as of 30 Jun 2017
%========================================================
 \enlargethispage{20pt}
 \bibitem{Bergqvist:2004}
  G.~Bergqvist and P.~Lankinen,
  ``Algebraic and differential Rainich conditions for symmetric trace-free tensors of higher rank'',
  Proc.\ Roy.\ Soc.\ Lond.\ A {\bf 461} (2005) 2181
  doi:10.1098/rspa.2004.1411
  [gr-qc/0405004].
  %%CITATION = doi:10.1098/rspa.2004.1411;%%
  %2 citations counted in INSPIRE as of 30 Jun 2017
%========================================================

\bibitem{Plebanski:1994}
  J.~F.~Plebanski and M.~Przanowski,
  ``Duality transformations in electrodynamics'',\\
  Int.\ J.\ Theor.\ Phys.\  {\bf 33} (1994) 1535.
  doi:10.1007/BF00670696
  %%CITATION = doi:10.1007/BF00670696;%%
  %5 citations counted in INSPIRE as of 30 Jun 2017
%========================================================

\bibitem{Torre:2013}
  C.~G.~Torre,
  ``The spacetime geometry of a null electromagnetic field'',
  Classical and Quantum Gravity {\bf 31} (2014) 045022
  doi:10.1088/0264-9381/31/4/045022
  [arXiv:1308.2323 [gr-qc]].
  %%CITATION = doi:10.1088/0264-9381/31/4/045022;%%
  %5 citations counted in INSPIRE as of 30 Jun 2017
 %======================================================== 


\bibitem{Krongos:2015}  
  D.~S.~Krongos and C.~G.~Torre,
  ``Geometrization conditions for perfect fluids, scalar fields, and electromagnetic fields'',
  J.\ Math.\ Phys.\  {\bf 56} (2015) no.7,  072503
  doi:10.1063/1.4926952
  [arXiv:1503.06311 [gr-qc]].
  %%CITATION = doi:10.1063/1.4926952;%%
  %3 citations counted in INSPIRE as of 30 Jun 2017
%========================================================  


\bibitem{Santos:2016}  
  W.~Cordeiro dos Santos,
  ``Introduction to Einstein--Maxwell equations and the Rainich conditions'',
  arXiv:1606.08527 [gr-qc].
  %%CITATION = ARXIV:1606.08527;%%
  %1 citations counted in INSPIRE as of 30 Jun 2017
%========================================================

\bibitem{Balfagon:2007}
  A.~C.~Balfagon,
  ``Rainich theory applied to $m$-rank tensors in $n$-dimensions'',
  %Submitted to: Gen.Rel.Grav.
  [arXiv:0709.1041 [gr-qc]].
  %%CITATION = ARXIV:0709.1041;%%

%========================================================  
 
%========================================================
\bibitem{Petrov}
 S.~Hervik, V. Pravda, and A. Pravdov\'a, 
 ``Type III and II universal spacetimes'',\\
J. Phys. Conf. Ser. {\bf600} (2015) 012066;  ERE 2014; 
 doi:10.1088/1742-6596/600/1/012066
  


%========================================================
% MATRIX ANALYSIS
%========================================================
\bibitem{HJ1}
R.~Horn and C.~Johnson, \emph{Matrix analysis}, \\
(Cambridge University Press, Cambridge, 1990)
%========================================================

\bibitem{HJ2}
R.~Horn and C.~Johnson, \emph{Topics in matrix analysis}, \\
(Cambridge University Press, Cambridge, 1994)
%========================================================
  
  
%========================================================
% NO SYMMETRY
%========================================================
\bibitem{Koutras}
A.~Koutras and C.~McIntosh,
``A metric with no symmetries or invariants'',\\
Class. Quantum Grav. {\bf13} (1996) L47--L49. 

%========================================================
% VSI AND RELATED
%========================================================

  \bibitem{Wils}
  P.~Wils,
  ``Homogeneous and conformally Ricci-flat pure radiation fields'',\\
  Classical and Quantum Gravity {\bf6} (1989) 1243--1251.
  
\bibitem{Koutras:1992}
 A.~Koutras,
``A spacetime for which the Karlhede invariant classification requires the fourth covariant derivative of the Riemann tensor'',\\
Classical and Quantum Gravity {\bf9} (1992) L143--Ll45. 

\bibitem{Pravda:2002}
V.~Pravda, A.~Pravdova, A.~Coley, and R.~Milson,\\
``All spacetimes with vanishing curvature invariants'',\\
Classical and Quantum Gravity {\bf19} (2002) 6213--6236.

\bibitem{Page:2009}
D.~Page, 
``Nonvanishing local scalar invariants even in VSI spacetimes with all polynomial curvature scalar invariants vanishing'',\\
Class. Quantum Grav. {\bf26} (2009) 055016;  doi:10.1088/0264-9381/26/5/055016

%========================================================
% ROSEN
%========================================================

\bibitem{Cropp:2010}
  B.~Cropp and M.~Visser,
  ``General polarization modes for the Rosen gravitational wave'',
  Classical and Quantum Gravity {\bf 27} (2010) 165022
  doi:10.1088/0264-9381/27/16/165022
  [arXiv:1004.2734 [gr-qc]].
  %%CITATION = doi:10.1088/0264-9381/27/16/165022;%%
  %7 citations counted in INSPIRE as of 30 Jan 2018
  
  \enlargethispage{20pt}
  \bibitem{Cropp:2011}
  B.~Cropp and M.~Visser,
  ``Polarization modes for strong-field gravitational waves'',\\
  J.\ Phys.\ Conf.\ Ser.\  {\bf 314} (2011) 012073
  doi:10.1088/1742-6596/314/1/012073
  [arXiv:1011.5904 [gr-qc]].
  %%CITATION = doi:10.1088/1742-6596/314/1/012073;%%
  %4 citations counted in INSPIRE as of 30 Jan 2018
  
 \bibitem{Wald}
 R.~M.~Wald, \emph{Gravitation}, (University of Chicago Press, USA, 1984).

%========================================================
\end{thebibliography}
\end{document}